\documentclass[10pt]{article} 
\usepackage[preprint]{rlc}

\usepackage{amssymb}            
\usepackage{mathtools}          
\usepackage{mathrsfs}           
\usepackage{graphicx}           
\usepackage{subcaption}         
\usepackage[space]{grffile}     
\usepackage{url}                
\usepackage{soul}

\usepackage{amsthm}
\usepackage{comment}
\usepackage{algorithm}
\usepackage{algpseudocode}
\usepackage{subcaption}
\usepackage{mathrsfs}
\usepackage{booktabs}
\usepackage{graphicx}
\usepackage{tikz}
\usetikzlibrary{positioning,fit,calc,arrows.meta,backgrounds}

\newtheorem{remark}{Remark}

\DeclareRobustCommand{\indicator}{\text{\usefont{U}{bbold}{m}{n}1}}

\definecolor{rougePOLY}{RGB}{211, 1, 47}
\definecolor{orangePOLY}{RGB}{255, 96, 5}
\definecolor{vertPOLY}{RGB}{1, 234, 83}
\definecolor{bleuPOLY}{RGB}{80, 187, 230}
\definecolor{violetPOLY}{RGB}{68, 17, 81}
\definecolor{olivePOLY}{HTML}{79745C}

\title{Cooperation Without Shared Rewards via Reciprocal
Q-Augmentation}

\author{Francesco Morri  \\
    francesco.morri@polymtl.ca \\
    Polytechnique Montréal \\
    Montréal, Canada
    \And
    Adel Abusitta \\
    adel.abu-sitta@polymtl.ca\\
    Polytechnique Montréal \\
    Montréal, Canada
    \And
    Quentin Cappart \\
    quentin.cappart@polymtl.ca \\
    Polytechnique Montréal, Mila \\
    Montréal, Canada \\
    UCLouvain \\
    Louvain-la-Neuve, Belgium
    \And 
    Margarida Carvalho \\
    carvalho@iro.umontreal.ca \\
    Université de Montréal, Mila\\
    Montréal, Canada}

\begin{document}

\maketitle

\begin{abstract}
General-sum games are more challenging than fully cooperative or adversarial settings because agents may deceive or exploit one another. Achieving high social welfare requires sustained cooperation, while a single defection can drive agents toward a welfare suboptimal Nash equilibrium. Moreover, assuming access to other agents' rewards or internal states is often unrealistic and implicitly introduces additional cooperation into the setting. To address this issue, we introduce \textit{Reciprocal Q-Augmentation} (RQA), a method that uses a latent-variable opponent model to augment the Q-function with estimates of how one's actions affects other agents. RQA allows agents to learn both cooperative and defective policies using only their own Q-function. We show that, in social dilemmas, agents that all follow the cooperative policy of RQA learn to cooperate faster and reach higher levels of cooperation than fully cooperative algorithms. Moreover, a single RQA agent that switches appropriately between its cooperative and defecting policies can induce cooperation in agents trained with standard RL algorithms, which would otherwise end up defecting.
We evaluate our approach on the Iterated Prisoner's Dilemma, Iterated Traveler's Dilemma, Iterated Stag Hunt, and \texttt{CleanUp-Lite}. Our experiments show that the algorithms implementing RQA outperforms fully cooperative methods with shared rewards and enables cooperation with algorithms that otherwise fail to reach high social welfare.
\end{abstract}

\section{Introduction}
Cooperation between multi-agent reinforcement learning agents is a critical problem, especially if the end goal is to deploy such agents in realistic and complex settings \citep{dafoe_open_2020}. This problem becomes particularly challenging when full cooperation is not necessarily unilaterally optimal from the perspective of each individual agent, despite Pareto dominating the Nash equilibrium the agents would otherwise reach, and thus making every player better off.
These settings fall into the category of general-sum games, which generalize both common-reward (fully cooperative) games, where objectives are fully aligned, and two-player zero-sum games, where they are strictly opposed. Falling at neither extreme introduces an additional difficulty: an agent must account not only for how its actions affect its own return, but also for whether the behavior of the other agents makes cooperation beneficial or exploitable.


The specific settings under which we study this problem are sequential social dilemmas \citep{leibo_multi-agent_2017}. Social dilemmas capture situations in which individually rational behavior can lead to collectively inefficient outcomes: exploiting a cooperative player can be more rewarding than mutual cooperation, while mutual defection can be preferable to being exploited. At the same time, mutual cooperation yields the highest collective welfare. Sequential social dilemmas extend this tension to temporally extended interactions, where cooperation and defection correspond to policies or behavioral patterns rather than isolated one-shot actions \citep{leibo_multi-agent_2017}. An agent must therefore reason about whether the behavior of the other player warrants cooperation over an extended interaction, while remaining able to react when that cooperation is not reciprocated.

The tension between cooperation and exploitation is then at the core of the problem we consider: \textit{How can an agent learning only from its own reward sustain cooperation, resist exploitation, and induce cooperation in a co-player that would otherwise defect?} Limiting our agents to observe only their own reward makes this question particularly challenging and relevant to realistic applications. 

\paragraph{Mechanisms for Cooperation}
Several mechanisms have been proposed to support the emergence of cooperation in social dilemmas. One possibility is to allow agents to select whom they interact with \citep{anastassacos_partner_2020, leung_learning_2026}. In \citet{anastassacos_partner_2020}, agents can observe the recent actions of other players and select partners based on their interaction history, while \citet{leung_learning_2026} studies a setting in which direct experience is supplemented by more limited third-party observations. In both cases, partner selection allows agents to preferentially interact with cooperative players and avoid or retaliate against defectors, effectively learning a tit-for-tat strategy~\citep{axelrod_evolution_1981}.
A different direction modifies the incentives under which agents learn. For example, \citet{wang_evolving_2019} uses an evolutionary process to learn intrinsic motivations that encourage altruistic behavior, modifying the reward signal on which individual agents learn. More recently, \citet{demir_fairness_2026} studies cooperation under asymmetric incentives and adapts fairness-based intrinsic motivations to account for differences between agents.
For a broader review of methods for promoting cooperation in multi-agent reinforcement learning and social dilemmas, we refer to \citet{du_review_2023}.
In this work, we are interested in a complementary setting in which cooperation must emerge without selecting interaction partners or explicitly modifying the agents' rewards, and to do so, we have our agent model its opponent. We restrict ourselves to two-player games, the setting of the classical social dilemmas we consider and the one in which reciprocity is unambiguous, since a defection can be attributed to a single opponent and the punishment directed at it.
\paragraph{Opponent Modeling}
Deciding whether cooperation should be maintained naturally requires reasoning about the behavior of the other agent. Opponent modeling provides a general framework for incorporating such information into an agent's decision making. Previous work has, for example, learned latent representations of other agents from behavioral observations \citep{rabinowitz_machine_2018}, explicitly incorporated learned opponent models into reinforcement learning \citep{tian_regularized_2019}, and used variational latent-variable models to represent different opponent behaviors \citep{papoudakis_variational_2020}. Related approaches use latent opponent representations to adapt online to unknown or non-stationary opponents \citep{fu_greedy_2022}.
Opponent modeling is a means rather than an end in our work: we do not introduce a new general opponent-modeling methodology. Instead, we use a latent-variable model to learn cooperative and adversarial strategies, which can then be used to induce cooperation. Given the current state, a latent variable encoding information about the other player, and a candidate action of the controlled agent, the model predicts the action of the other player. This differs from simply identifying a fixed opponent policy: the conjecture is intended to model how the other player's behavior responds to our own actions.

\paragraph{Reciprocity and learning-aware adaptation}
Most closely related to our approach are methods that use reciprocity or learning-aware behavior to cooperate while limiting vulnerability to exploitation. \citet{lerer_maintaining_2018} generalize ideas from tit-for-tat to sequential social dilemmas by learning cooperative and defective behaviors and switching between them according to the behavior of the other player. We introduce a switching mechanism inspired by this idea, but that does not require learning policies for both players in advance. Instead, we train the cooperative and adversarial policies concurrently, and choose between them based on the predicted response of the other agent. Moreover, our alternative to cooperation is an explicitly adversarial policy, rather than a standard self-interested policy trained independently.

\citet{eccles_learning_2019} build on reciprocity, with learning agents that adapt their social behavior to that of their co-players. Their approach does so through an intrinsic motivation mechanism that modifies the learner's effective reward according to an estimate of the other player's sociality. In contrast, our approach does not modify the reward function, but explicitly chooses between cooperative and adversarial behaviors based on a model of the opponent.

Another closely related objective is that of ``foolproof'' cooperation \citep{jacq_foolproof_2020}, which seeks cooperative strategies that cannot profitably be exploited by selfish agents. Agents can fully communicate and access each others actions and rewards, while using a jointly obtained cooperative Q-function to assess the level of cooperation. The resulting algorithm mostly prevents exploitation when used against a standard RL algorithm, being able to induce cooperation only when multiple instances are deployed in the same game.

Finally, \citet{meulemans_multi-agent_2025} study how learning-aware agents can achieve cooperation by accounting for and shaping the learning dynamics of other agents, including in sequential social dilemmas. This provides a complementary mechanism for adapting cooperation to other learners. Our approach does not explicitly model the opponent's learning rule or its parameter updates; instead, it uses the opponent's observed behavior to infer a latent behavioral state and predict its response to the agent's own actions.

\paragraph{Contributions}
Taken together, these approaches demonstrate that cooperation can be supported through reciprocity, incentive modification, opponent adaptation, or reasoning about other learners. Our goal is to retain the ability to cooperate conditionally, cooperating when cooperation is reciprocated and responding adversarially otherwise, while requiring neither shared nor modified rewards, explicit communication, nor privileged access to the other agent's reward or policy. At the same time, both behaviors and the mechanism used to select between them can be learned during interaction. To achieve these goals, we introduce the following contribution:
(1) An innovative method for learning cooperative behavior in general-sum games using only individual rewards, without requiring shared rewards, explicit communication, or access to the other agents' policies. (2) We show that an agent trained with our method can be frozen and used as a \textit{cooperation teacher}, inducing cooperative behavior in independently learning agents that would otherwise converge to non-cooperative behavior; (3) We further show that this teaching effect does not require a separately converged self-play phase: cooperation can be induced while the proposed agent and its co-player are still learning concurrently. In the considered sequential social dilemmas, the resulting agents achieve social welfare and cooperation levels comparable to, and in some settings better than, fully cooperative baselines. 


\section{Settings}
The environments we consider fall into the category of stochastic games \citep{kuhn_6_2020}, consisting of the tuple $(\mathcal{N}, \mathcal{S}, $ $\{A_i\}_{i\in\mathcal{N}}, \{R_i\}_{i\in\mathcal{N}}, P_T, \nu, T)$: $\mathcal{N}$ is the set of $N$ players, $\mathcal{S}$ is the set of states, $A_i$ is the set of actions $a_i\in A_i$ for each agent $i$, with $A=A_1\times\cdots\times A_N$, $R_i:A\times\mathcal{S}\to\mathbb{R}$ is the reward function of agent $i$, and $r_i$ the value of the reward,
finally, $P_T:A\times\mathcal{S}\to\Delta(\mathcal{S})$ is the transition function.
The notation $\Delta(\mathcal{X})$ represents the set of probability distributions over the set $\mathcal{X}$. The parameters $\nu$ and $T$ represent the discount factor and the game horizon, which is finite. Notice that we omit the observations for the players, since in our settings every player observes the full state~$s$.
Each agent aims to learn a policy $\pi_i$, with $\pi=(\pi_1,\cdots,\pi_N)$, that maximizes its discounted expected return: 
$$J_i(\pi_i, \pi_{-i}) = \mathbb{E}_{s\sim P_T,a\sim \pi}\left[\sum_{t=1}^T\nu^tR_i(s^t,a^t)\right], \forall i \in \{1,\dots, N\}$$ where $\pi_{-i}$ represents the policies of all other players except $i$. 
We will consider algorithms explicitly learning a Q-function that receives the joint action $a$ of all agents, which is the standard centralized-training-decentralized-execution paradigm \citep{lowe_multi-agent_2017}. Further, we consider the case of $N=2$, introducing our algorithm under this assumption, and leaving a scalable generalization to $N>2$ players as future work.

\section{Reciprocal Q-Augmentation}
With our approach we set to achieve two main goals: 
i) develop an agent that is a very good cooperator, in that it \textit{pushes} towards cooperation; ii) the same agent should not be easily exploited by free-riders. To achieve this goal, we propose a method called \textit{reciprocal Q-augmentation} (RQA). We present it incrementally: we first show how an agent can learn to cooperate while observing only its own reward, and then introduce a mechanism that prevents it from being exploited while still promoting cooperation.

Our core hypothesis is that maintaining an internal model of the other player is fundamental for learning when to cooperate and when to defect, especially when agents do not have access to each other's rewards. Building on this idea, our method starts by learning an opponent model, which is then used to augment the agent's Q-function. This augmentation encourages cooperation while also preventing the agent from being exploited. In Section~\ref{subsec:opponent_model}, we describe how we model the other players in practice. To do so, we import into multi-agent reinforcement learning the game-theoretic notion of a \textit{conjecture}: rather than predicting the opponent's behavior alone, we predict how the opponent responds to the agent's own action. In Section~\ref{subsec:coop_aug}, we then show how this model is incorporated into the Q-function, in Section~\ref{subsec:switching} we introduce how the agent can avoid exploitation and improve cooperation, finally in Section~\ref{subsec:rqa-masac}. In the following sections, we adopt the perspective of agent~$i$ facing its opponent~$j$.

\subsection{Latent Opponent Modeling}
\label{subsec:opponent_model}
\begin{figure}[bt]
\centering
\resizebox{0.8\columnwidth}{!}{%
\begin{tikzpicture}[
    x=0.61cm,
    y=0.72cm,
    >=Latex,
    line width=0.7pt,
    var/.style={
        draw,
        rounded corners=1.5pt,
        minimum width=0.78cm,
        minimum height=0.52cm,
        inner sep=1pt,
        fill=blue!20,
        font=\small
    },
    latent/.style={
        draw,
        circle,
        minimum size=0.72cm,
        inner sep=0.5pt,
        fill=blue!20,
        font=\small
    },
    arr/.style={->, thick},
    conn/.style={thick}
]


\filldraw[
    fill=green!25,
    draw=green!50!black,
    rounded corners=4pt
]
(-0.8,-0.9) rectangle (5.5,1.45);

\filldraw[
    fill=cyan!22,
    draw=cyan!50!black,
    rounded corners=4pt
]
(-0.9,-4.25) rectangle (8.85,-2.1);

\filldraw[
    fill=yellow!35,
    draw=yellow!55!black,
    rounded corners=4pt
]
(9.05,-4.25) rectangle (11.6,1.2);

\node[var] (ri) at (0,0.35) {$r_i$};
\node[var] (s)  at (1.6,0.35) {$s$};
\node[var] (ai) at (3.2,0.35) {$a_i$};
\node[var] (aj) at (4.8,0.35) {$a_j$};

\node[font=\bfseries\small]
    at (0.4,-0.42) {ENV};

\node[latent] (f) at (0,-3.15) {$f_i^j$};

\node[var] (h) at (2.1,-3.15)
    {$h_{i,j}$};

\node[latent] (q) at (4.1,-3.15)
    {$q_i^j$};

\node[var] (z) at (6.1,-3.15)
    {$z_{i,j}$};

\node[latent] (gamma) at (8.0,-3.15)
    {$\gamma_i^j$};

\node[font=\bfseries\small]
    at (3.5,-2.38) {OPPONENT MODEL};

\node[var] (atilde) at (8.0,0)
    {$\tilde a_j$};

\node[latent] (Q) at (10.35,0)
    {$Q_i^{\mathrm{C/D}}$};

\node[latent] (pi) at (10.35,-2.25)
    {$\pi_i^{\mathrm{C/D}}$};

\node[
    font=\bfseries\small,
    align=center
]
at (10.35,-3.65) {FINAL\\POLICY};

\draw[arr] (f) -- (h);
\draw[arr] (h) -- (q);
\draw[arr] (q) -- (z);
\draw[arr] (z) -- (gamma);

\draw[arr]
    (h.south) -- (2.1,-4.15) -- (0,-4.15) -- (f.south);

\draw[conn] (1.6,-0.03) -- (1.6,-1.55);
\draw[conn] (3.2,-0.03) -- (3.2,-1.55);
\draw[conn] (4.8,-0.03) -- (4.8,-1.55);

\draw[conn] (0,-1.55) -- (4.8,-1.55);

\draw[arr] (0,-1.55) -- (f.north);

\draw[conn] (1.6,-0.03) -- (1.6,-1.90);
\draw[conn] (3.2,-0.03) -- (3.2,-1.90);
\draw[conn] (4.8,-0.03) -- (4.8,-1.90);

\draw[conn] (1.6,-1.90) -- (7.75,-1.90);

\draw[arr] (7.75,-1.90) -- (gamma.north);

\draw[arr]
    (gamma.north) -- (atilde.south);

\draw[arr]
    (atilde.east) -- (Q.west);

\draw[arr]
    (Q) -- (pi);

\draw[conn]
    (0,0.73) -- (0,1.85) -- (10.35,1.85);

\draw[conn] (1.6,0.73) -- (1.6,1.85);
\draw[conn] (3.2,0.73) -- (3.2,1.85);
\draw[conn] (4.8,0.73) -- (4.8,1.85);

\draw[arr]
    (10.35,1.85) -- (Q.north);

\end{tikzpicture}
}
\caption{Schematic of the different modules of the RQA method.}
     \label{fig:algorithm}
\end{figure}
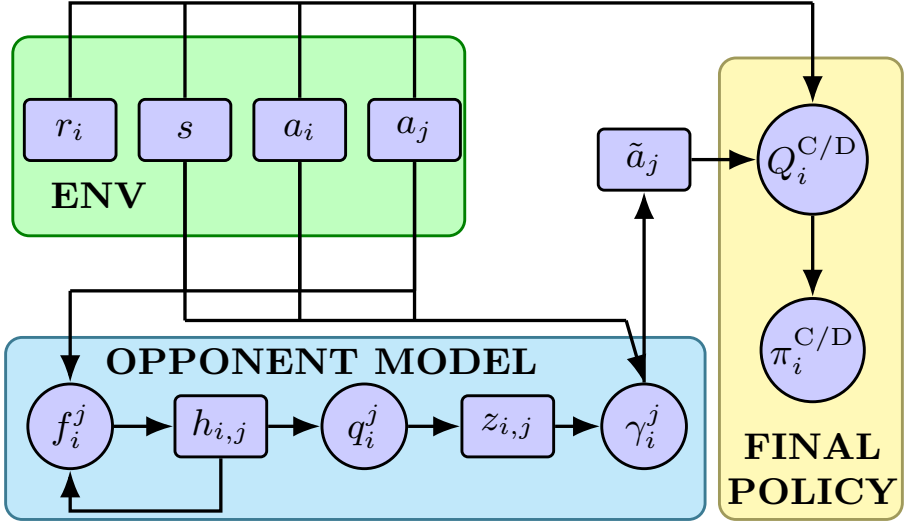

The structure of our opponent model builds on \citet{fu_greedy_2022}, with a simplified version of their latent-variable architecture. We do not claim a novel opponent-modeling approach: the opponent model is only a tool to augment the Q-function, and other models could be used instead. Our approach only requires the model to predict the response of player~$j$ to a candidate action of player~$i$, rather than the behavior of $j$ alone. Such predictions are known in game theory as \textit{conjectures} \citep{figuieres_theory_2004, morri_steering_2025}, and they are what allows the opponent model to be integrated into the augmented Q-function.
Our model is composed of three modules: an \textit{encoder}, an \textit{inference function}, and a \textit{conjecture}. We provide a high-level overview of this model in the blue box named \textit{opponent model} in Fig.~\ref{fig:algorithm}. The green box groups the elements provided by the environment, namely the state, the reward, and the actions of both players.

\paragraph{Encoder Function.} The first component of the model is the encoder, whose role is to produce a single variable \textit{encoding} the current information about the agent $j$ given the state and action of agent $i$. We define it as
\begin{equation}
f_i^j(h_{i,j},s,a_i,a_j;\psi_i^j):\mathcal{H}\times \mathcal{S}\times A_i\times A_j\to\mathcal{H},
\end{equation}
where $\mathcal{H}$ is the space of hidden states. It receives the state $s$ of the environment, both actions $a_i,a_j$, and the hidden state variable $h_{i,j}$, obtained recursively at each step $t$ as $h_{i,j}^t=f_i^j(h_{i,j}^{t-1},s^{t-1},a_i^{t-1},a_j^{t-1};\psi_i^j)$. It is parameterized by~$\psi_i^j$.

\paragraph{Inference Function.}
A latent variable $z_{i,j}^t$ is then obtained from the hidden state $h_{i,j}^t$ through the inference function $q_i^j(z_{i,j}^t \mid h_{i,j}^t; \phi_i^j)$, which we parameterize as a Gaussian distribution:
\begin{equation}
    q_i^j(z_{i,j}^t \mid h_{i,j}^t; \phi_i^j) = \mathcal{N}\big(\mu_{i,j}(h_{i,j}^t),\, \Sigma_{i,j}(h_{i,j}^t)\big),
\end{equation}
where the mean $\mu_{i,j}$ and the covariance $\Sigma_{i,j}$ are functions of $h_{i,j}^t$ learned with parameters $\phi_i^j$. We directly learn a diagonal form for $\Sigma_{i,j}$, further when we refer to these two quantities without specifying the arguments, we explicitly show the dependence on the time step $t$, writing $\mu_{i,j}^t,\Sigma_{i,j}^t$.


\paragraph{Conjecture Function.}
Finally, given the latent variable $z_{i,j}$ sampled from the inference function, player~$i$ predicts how player~$j$ would respond to each of its candidate actions. This prediction is given by the conjecture function
\begin{equation}
    \gamma_i^j(a_i, s, z_{i,j}; \theta_i^j) : A_i \times \mathcal{S} \times \mathcal{Z} \to \Delta(A_j),
\end{equation}
where $\mathcal{Z}$ is the latent space and $\theta_i^j$ are the parameters of the conjecture. It takes as input a candidate action $a_i$ of player~$i$, the state $s$, and the latent variable $z_{i,j}$, and outputs a probability distribution over the actions of player~$j$.

\paragraph{Loss Definition.} 
Given an input composed of $(a_i^t, a_j^t, s^t, z_{i,j}^t, \allowbreak z_{i,j}^{t-1},\allowbreak \mu_{i,j}^t,\Sigma_{i,j}^t)$, the three modules are trained together using the following loss:
\begin{equation}
    \begin{split}
        \ell_{i,j}^t &= -\log\gamma_i^j(a_j^t|a_i^t,s^t,z_{i,j}^t) +\\
        &+\beta_i D_{KL}\left(\mathcal{N}(\mu_{i,j}^t,\operatorname{diag}(\Sigma_{i,j}^t))||\mathcal{N}(z_{i,j}^{t-1},\sigma_i^2\indicator)\right)
    \end{split}
    \label{eq:latent_loss}
\end{equation}
The first term is the \textit{prediction loss},
which quantifies how good our current conjecture is. The second term, weighted by the hyperparameter $\beta_i$ is the \textit{temporal loss}, which limits the inference model from changing too rapidly from one iteration to the next, measuring its distance from a normal distribution centered in the previous latent variable and with standard deviation $\sigma_i$, the other hyperparameter of the loss.

\paragraph{Training Algorithm.} 
We train this model by stochastic gradient descent on Eq.~\eqref{eq:latent_loss}, averaged over $M$ contiguous trajectory fragments sampled from a replay buffer rather than independent transitions. Each fragment $x_m$, $m\in[0,M]$, provides a temporally ordered sequence of length $\tau$ of actions of the other player, which allows the model to recognize patterns in its strategy.
Once the loss from Eq.~\eqref{eq:latent_loss} is computed over all fragments, the final update is done with:
\begin{equation}
    \mathcal{L}_{i,j} = \frac{1}{M\tau}\sum_{m=0}^M\sum_{k=t_m}^{t_m+\tau}\ell_{i,j}^k(x_m),
\end{equation}
as reported in Algorithm~\ref{alg:opp_model_training}.


\begin{algorithm}[h]
\caption{Latent Opponent Model Update.}
\label{alg:opp_model_training}
\begin{algorithmic}[1] 
\State Sample $M$ trajectories fragments $\left\{[a_i^k,a_j^k,s^k]_{k=t_m}^{t_m+\tau}\right\}_{m=0}^M$, $\mathcal{L}_{i,j}\gets0$
\For{each fragment $m=[0,\dots,M]$}
\State $h_{i,j}^{t_m}\gets0,z_{i,j}^{t_m}\gets0,\ell_{i,j}^m\gets0$
\For{each actions point $k=[t_m+1,\dots,t_m+\tau]$}
\State $h_{i,j}^k\gets f_i^j(h_{i,j}^{k-1},a_i^{k-1},a_j^{k-1},s^{k-1})$
\State $(\mu_{i,j}^k,\Sigma_{i,j}^k)\gets q_i^j(h_{i,j}^k)$
\State $z_{i,j}^k \sim\mathcal{N}(\mu_{i,j}^k,\operatorname{diag}(\Sigma_{i,j}^k))$ 
\State $p_{i,j}^k\gets \log \gamma_i^j(a_j^k|a_i^k,s^k,z_{i,j}^k)$
\State $d_{i,j}^k\gets D_{KL}\left(\mathcal{N}(\mu_{i,j}^k,\operatorname{diag}(\Sigma_{i,j}^k))||\mathcal{N}(z_{i,j}^{k-1},\sigma_i^2\indicator)\right)$
\State $\ell^m_{i,j}\gets \ell^m_{i,j} - p + \beta_i d$
\EndFor
\State $\mathcal{L}_{i,j}\gets\mathcal{L}_{i,j}+\frac{1}{\tau}\ell^m_{i,j}$
\EndFor
\State Compute $\nabla \frac{1}{M}\mathcal{L}_{i,j}$
\State Optimizer step for $f_i^j,q_i^j,\gamma_i^j$
\end{algorithmic}
\end{algorithm}
\subsection{Cooperative Q-Augmentation}
\label{subsec:coop_aug}
Our main contribution is on how to use the opponent model to augment the Q-function. This is loosely inspired by the idea of \textit{perspective taking}, which is a phenomenon mostly studied and observed in behavioral economics and psychology, showing how  considering the perspective of your opponent (or teammate) can alter your decision, both towards a more cooperative action \citep{galinsky_why_2008}, or a more egotistical one \citep{epley_when_2006}.
Following this idea we introduce a cooperative augmentation for the Q-function:
\begin{equation}
    Q_i^C(s,a_i,a_j) = Q_i(s,a_i,a_j) + \alpha_i^C Q_i(s,\gamma_i^j(a_i,s,z_{i,j}),a_i).
    \label{eq:cooperative_Q}
\end{equation}
The second term, weighted by the hyper-parameter $\alpha_i^C\in[0,\infty]$, is what pushes the agent towards cooperation.
Agent $i$ is effectively trying to maximize the reward from the perspective of agent $j$, in addition to its own. In this sense, $Q_i^C$ can be seen as a crude approximation of the cumulative expected reward of the two agents, which we consider to be the social welfare of the system, and agent $i$ can be seen as maximizing this approximated social welfare.

\begin{remark}
    From Eq.~\eqref{eq:cooperative_Q}, one might expect that a perfectly symmetric game is needed to obtain a useful policy. In practice, we only require the two agents to have the same available actions and the same game \textit{logic}. For example, in the Prisoner's Dilemma, it is sufficient that the payoff order exploiting $>$ mutual cooperation $>$ mutual defection $>$ exploited holds for both players, even with different numerical rewards. In Section~\ref{sec:experiments}, we validate this on asymmetric games.
\end{remark}

The benefit of Eq.~\eqref{eq:cooperative_Q} is that it does not require introducing a second Q-function, sharing rewards between agents, or accessing the other agent's policy. Moreover, because the modification is applied directly to the Q-function, independently of the policy learning algorithm, it can be incorporated into any actor-critic method that learns a Q-function, such as Multi-Agent Deep Deterministic Policy Gradient (MADDPG) \citep{lowe_multi-agent_2017} or Multi-Agent Soft Actor Critic (MASAC) \citep{haarnoja_soft_2018}. Finally, the modification is agnostic to the action-space type and can therefore be applied to both discrete and continuous environments.
The augmented Q-function is represented in the yellow box (\textit{final policy}) in Fig.~\ref{fig:algorithm}, schematically showing how the environment and the opponent model connect to it.

\subsection{Enabling Reciprocal Cooperation via Cooperative and Defective Q-Functions}
\label{subsec:switching}
As mentioned in the previous section, considering the opponent's perspective can also lead to adversarial and/or selfish behavior. Using this intuition we can easily obtain an augmented Q-function that can be used to learn a defective policy:
\begin{equation}
    Q_i^D(s,a_i,a_j) = Q_i(s,a_i,a_j) - \alpha_i^DQ_i(s,\gamma_i^j(a_i,s,z_{i,j}),a_i).
    \label{eq:defective_Q}
\end{equation}
The only difference between Eq.~\eqref{eq:cooperative_Q} and Eq.~\eqref{eq:defective_Q} is the sign in front of the modified Q-function. We also introduce a second parameter $\alpha_i^D$, since we may want to give different weights to the cooperative and defective case. We represent schematically how all the modules of our algorithm fit together in Fig.~\ref{fig:algorithm}.

With Eqs.~\eqref{eq:cooperative_Q}-\eqref{eq:defective_Q}, our agent can learn concurrently two policies, $\pi^C,\pi^D$ (see the yellow box in Fig.~\ref{fig:algorithm}). The goal for the agent is  to understand when to switch between the cooperative and defective policy, timing it so that whoever is the opponent is incentivized to cooperate. The switching logic we introduce is a streamlined version of what \citet{lerer_maintaining_2018} propose, inspired by generalizing the classical tit-for-tat strategy \citep{axelrod_evolution_1981}. In addition our technique does not require a pre-training session to obtain the policies for both players, or sharing the rewards, and the defective policy is not just a standard single agent RL policy.
The agent tracks a one-sided CUSUM statistics~\citep{page_continuous_1954} of the \textit{cooperation level} $G_t$, updated at each step as
\begin{equation}
    \begin{aligned}
        G_t &= \max\{0,\, G_{t-1} + \delta_t\},\\
        \delta_t &= \sum_{\bar{a}_j \in A_j} \pi_i^C(\bar{a}_j \mid s^t)\, Q_i^C(s_i^t, a_i^t, \bar{a}_j) - Q_i^C(s_i^t, a_i^t, a_j^t).
    \end{aligned}
\end{equation}
The term $\delta_t$ measures how much less value the action actually played by $j$ brings to $i$, compared to what is expected under $\pi_i^C$. If $G_t$ exceeds a threshold $\hat{G}$, the agent switches to its defecting policy for $K$ steps, sets $G_t$ to 0 and does not evaluate $\delta$. After the $K$ steps, the agent returns to the cooperative policy. The overall behavior is controlled by the hyperparameters $K$ and $\hat{G}$.
Since agent $i$ has no access to $\pi_j$ to compute $\delta_t$, it uses $\pi_i^C$ as a proxy for the ideal cooperative opponent behavior, consistent with the perspective-taking assumption of Eq.~\eqref{eq:cooperative_Q}. Further, we explicitly sum over all actions, which is doable only on discrete action spaces. In the case of continuous actions, we would replace this sum by Monte Carlo sampling.
As we mentioned, the intuition for this switching logic is to follow a tit-for-tat like behavior: start cooperating, switch to defection if opponent is observed defecting, return to cooperation after a punishing period.

\subsection{RQA-MASAC}
\label{subsec:rqa-masac}
With Eqs.~\eqref{eq:cooperative_Q}~-~\eqref{eq:defective_Q} we have a methodology to enrich the Q-function for cooperation (or defection). Using these results we can improve an existing MARL algorithm (MASAC), and we will refer to this modified version as RQA-MASAC. Specifically we show how the augmented Q-function affects the training of the policy. The critic learning part remains the same as for the vanilla configuration, so we do not discuss that here, we just recall that a key feature of the algorithm is having two critics, which in the following equations we index with $k$.
Our modification becomes relevant when considering the policy loss.
When $A_i$ is discrete, the policy $\pi_i(a_i\mid s)$ is a categorical distribution over the actions parameterized by $\omega_i$. Let us assume for now that the action spaces of all agents are small enough that we can fully enumerate them without issues.
We define:
\begin{equation*}
    q_{i,k}(s,a_i) = \sum_{a_j}\pi_j(a_j|s) Q_{i,k}(s,a_i,a_j),
\end{equation*}
and we consider the cooperative modification:
\begin{equation}
    \Bar{q}_i^{C}(s,a_i) = \min_{k=1,2}q_{i,k}(s,a_i)+\alpha_i^CQ_{i,k}(s,\gamma_i^j(a_i,s,z_{i,j}),a_i),
    \label{eq:masac_discrete_barq}
\end{equation}
where the min is taken over the two critics characteristic of the soft-actor critic algorithm.
The loss is:
\begin{equation}
    J_i^C(\omega_i) = \sum_{a_i}\pi_i(a_i\mid s)\left[\alpha_i\log\pi_i(a_i\mid s)-\Bar{q}_i^C\right],
    \label{eq:masac_discrete_policy_loss}
\end{equation}
and we want to obtain $\nabla_{\omega_i}J_i^C(\omega_i)$, which we can directly derive (we omit the arguments of the functions for clarity):
\begin{subequations}
\begin{align}
    \nabla_{\omega_i}J_i^C(\omega_i) & = \sum_{a_i} \nabla_{\omega_i}\pi_i\cdot \alpha_i\log\pi_i + \pi_i\alpha_i \nabla_{\omega_i}\log\pi_i-\nabla_{\omega_i}\pi_i\Bar{q}_i^C\\
    & = \sum_{a_i} \nabla_{\omega_i}\pi_i\cdot \alpha_i\log\pi_i + \alpha_i \nabla_{\omega_i}\pi_i-\nabla_{\omega_i}\pi_i\Bar{q}_i^C\\
    & = \sum_{a_i}\nabla_{\omega_i}\pi_i\left[\alpha_i\log\pi_i+\alpha_i-\Bar{q}_i^C\right],
\end{align}
\end{subequations}
where between the first and second line we used the policy gradient trick.
Considering then that $\sum_{a_i}\nabla_{\omega_i}\pi_i\alpha_i= \alpha_i\nabla_{\omega_i}\sum_{a_i}\pi_i =  0$, and writing explicitly Eq.~\eqref{eq:masac_discrete_barq}, we get:
\begin{equation}
\begin{split}
     \nabla_{\omega_i}J_i^C(\omega_i) = \sum_{a_i} \nabla_{\omega_i} \pi_i(a_i\mid o_i) &\Big[ \alpha_i \log\pi_i(a_i\mid o_i) - \Bar{q}_i(s,a_i)+\\
     &- \alpha_i^C \Bar{Q}_i(s,\gamma_i^j(a_i),a_i) \Big],
\end{split}
    \label{eq:masac_discrete_cooperative_loss}
\end{equation}
where $\Bar{q}$ and $\Bar{Q}$ represent the min between the two critics.\\
In the defective case, we only need to switch a sign:
\begin{equation}
    \begin{split}
     \nabla_{\omega_i}J_i^D(\omega_i) = \sum_{a_i} \nabla_{\omega_i} \pi_i(a_i\mid o_i) &\Big[ \alpha_i \log\pi_i(a_i\mid o_i) - \Bar{q}_i(s,a_i)+\\
     &+ \alpha_i^D \bar{Q}_i(s,\gamma_i^j(a_i),a_i) \Big].   
\end{split}
    \label{eq:masac_discrete_defective_loss}
\end{equation}
Our modification effectively adds a term to the policy gradient that steers the policy toward either increasing or decreasing the expected reward from the other agent’s perspective.

\section{Numerical Experiments}
\label{sec:experiments}
For our experiments, we benchmark RQA-MASAC against ISAC \citep{haarnoja_soft_2018}, MASAC \citep{haarnoja_soft_2018}, IPPO \citep{witt_is_2020}, MAPPO \citep{yu_surprising_2022}, IQL \citep{ganapathi_subramanian_partially_2021}, and QMIX \citep{rashid_qmix_2018}. In particular QMIX is used as a cooperative baseline, since it explicitly considers the cumulative reward of both agents.
For the different algorithms, we consider their implementations in BenchMARL \citep{bettini_benchmarl_2024}, adding our modification on top of that for MASAC, while the different environments we test are implemented using PettingZoo \citep{terry_pettingzoo_2021}.
All algorithms were fine-tuned on each environment separately, searching for parameters maximizing the cumulative reward. The full code to run the experiments with specific details about the implementation and the hardware used can be found in the online repository \url{https://anonymous.4open.science/r/reciprocal-Q-augmentation-9251/}.
\begin{figure*}[tb]
    \centering
    \begin{subfigure}[t]{0.32\textwidth}
        \includegraphics[width=\linewidth]{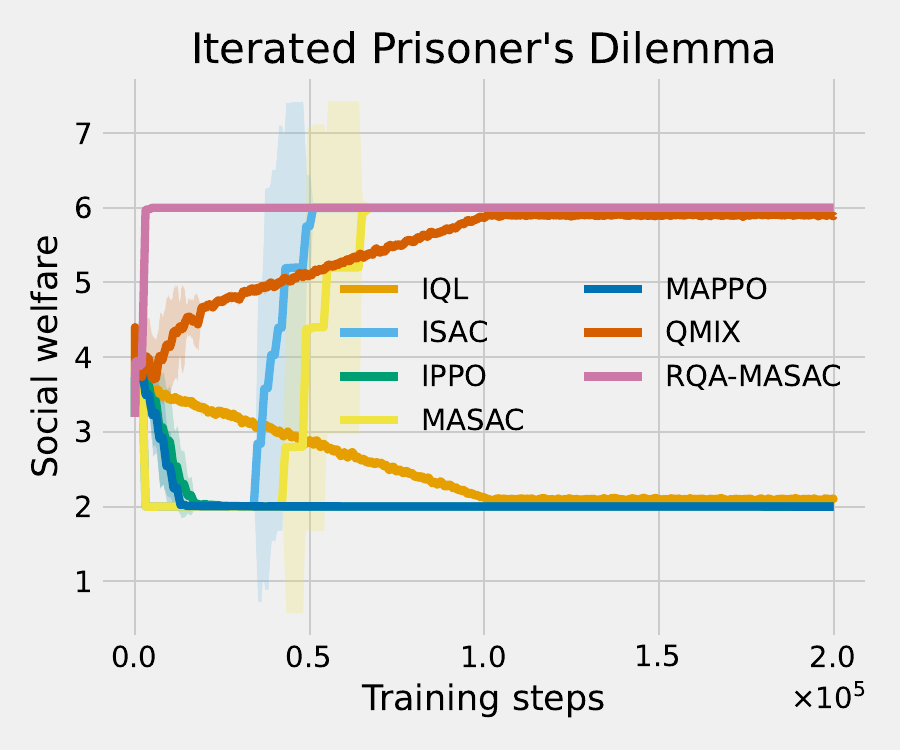}
        \caption{}
        \label{fig:ipd_welfare}
    \end{subfigure}
    \hfill
    \begin{subfigure}[t]{0.32\textwidth}
        \includegraphics[width=\linewidth]{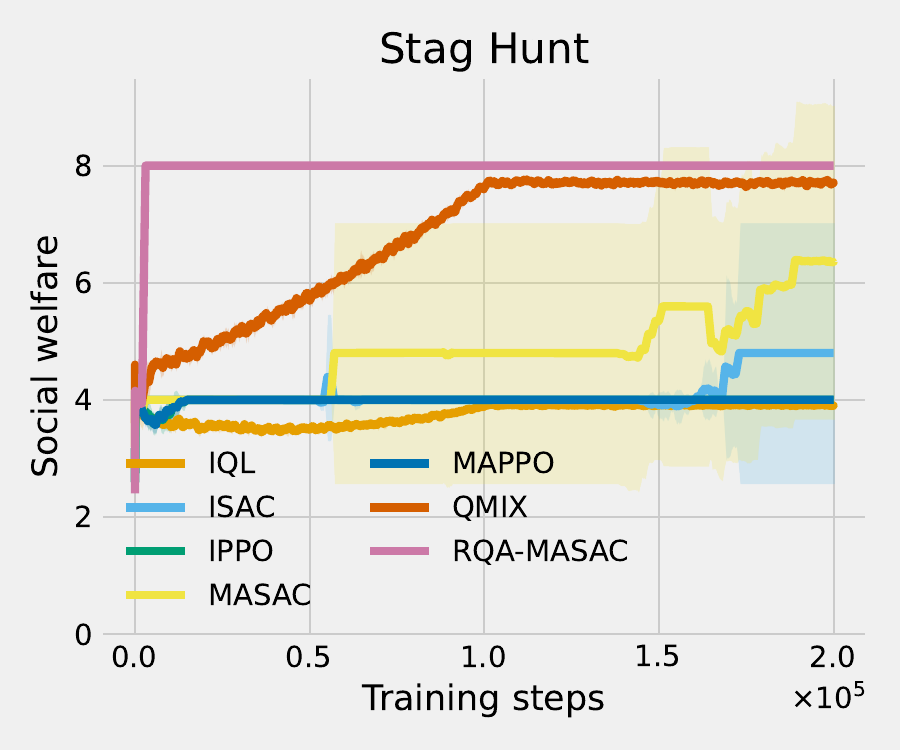}
        \caption{}
        \label{fig:stag_welfare}
    \end{subfigure}
    \hfill
    \begin{subfigure}[t]{0.32\textwidth}
        \includegraphics[width=\linewidth]{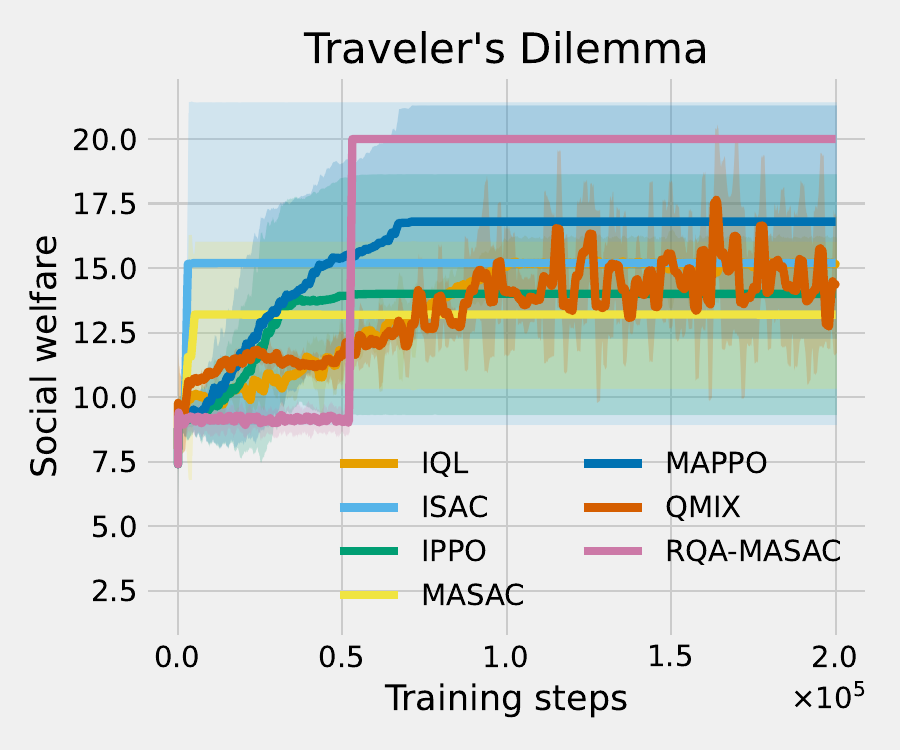}
        \caption{}
        \label{fig:trav_welfare}
    \end{subfigure}
    
    \begin{subfigure}[t]{0.32\textwidth}
        \includegraphics[width=\linewidth]{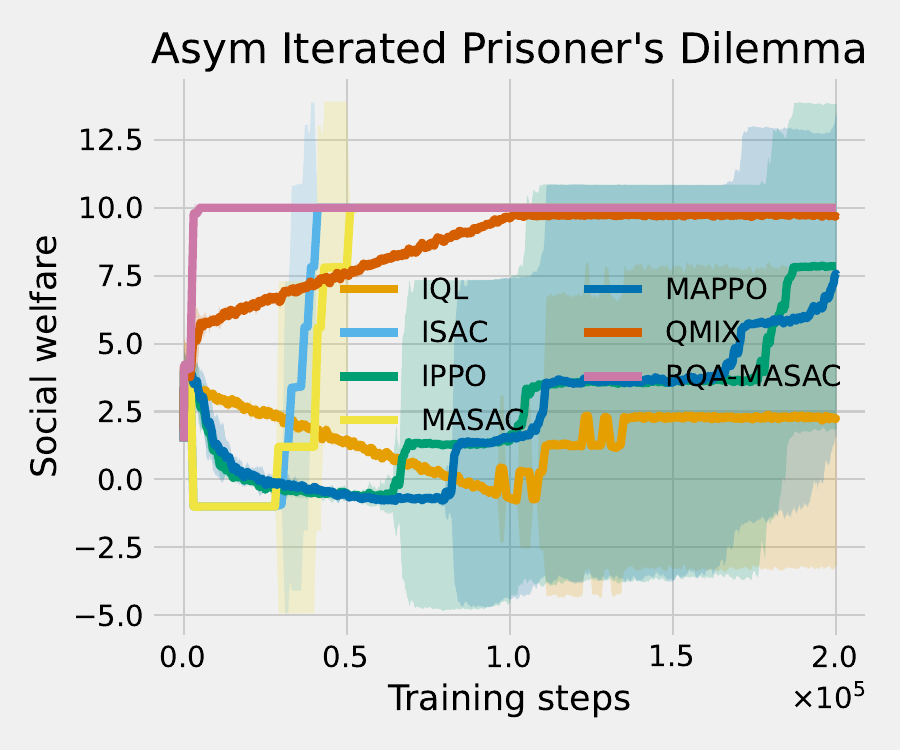}
        \caption{}
        \label{fig:asym_ipd_welfare}
    \end{subfigure}
    \hfill
    \begin{subfigure}[t]{0.32\textwidth}
        \includegraphics[width=\linewidth]{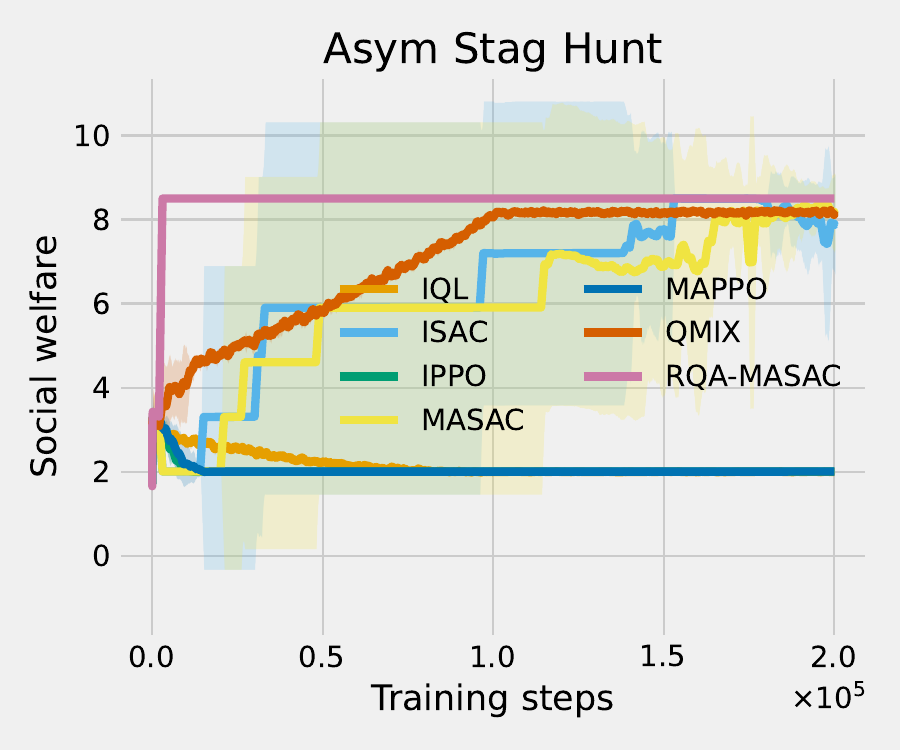}
        \caption{}
        \label{fig:asym_stag_welfare}
    \end{subfigure}
    \hfill
    \begin{subfigure}[t]{0.32\textwidth}
        \includegraphics[width=\linewidth]{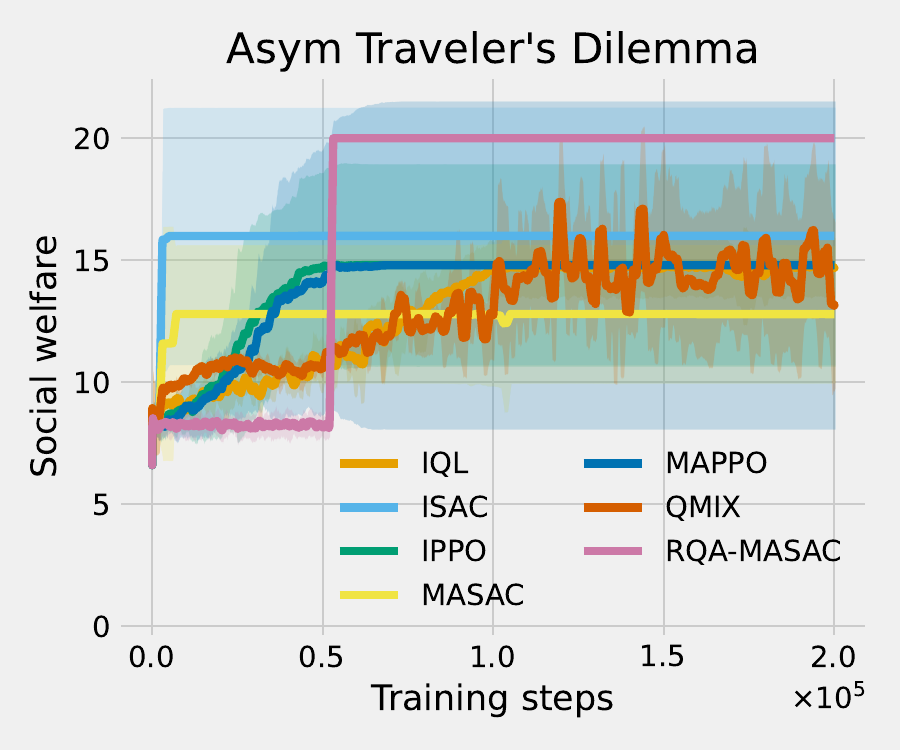}
        \caption{}
        \label{fig:asym_trav_welfare}
    \end{subfigure}
    \caption{Social welfare for the matrix games, averaged average over 5 different seeds and smoothed over 1000 training steps; the shaded area represents the $95\%$ confidence interval. (a-b-c): the three symmetric matrix games. In all three, RQA-MASAC converges to the optimal solution and maintains it, while the other algorithms struggle. (d-e-f): same games with asymmetric payoffs, where RQA-MASAC again converges consistently to the optimum, beating the other algorithms.}
    \label{fig:matrix_games_welfare}
\end{figure*}

\subsection{Environments}
We test our algorithm first on matrix games and then on a more challenging task: \texttt{CleanUp-Lite}.
\paragraph{Matrix Games}
We consider three classical matrix games: the Prisoner's Dilemma \citep{tucker1950twoperson, axelrod_evolution_1981}, Stag Hunt \citep{Skyrms_2003}, and the Traveler's Dilemma \citep{basu_travelers_1994}, with each respective payoff reported in Tables~\ref{tab:ipd}-\ref{tab:ish}-\ref{tab:itd}. The payoff for the Traveler's Dilemma is computed as follows:
\begin{equation*}
    r_1 =
    \begin{cases}
        & a_1 \mbox{ if } a_1=a_2\\
        & a_1+2 \mbox{ if } a_1<a_2\\
        & a_2 - 2 \mbox{ if } a_1>a_2
    \end{cases},\quad 2\leq a_1,a_2\leq 10
\end{equation*}
The Nash equilibrium for the Prisoner's Dilemma is mutual defection $(D,D)$, while the optimum social welfare is mutual cooperation $(C,C)$; Stag Hunt has two pure Nash: hunting together $(H,H)$ and foraging together $(F,F)$, of the two the optimum is hunting together; finally the Traveler's Dilemma has a single Nash, $(2,2)$, while the social optimum is $(10,10)$. The two dilemmas are used to showcase the risk of exploitation: cooperating with a defecting opponent in the Prisoner's Dilemma, and playing a high number with an opponent choosing a lower one for the Traveler's Dilemma; while Stag Hunt requires coordination: hunting together is better for both players, but hunting alone gives a negative reward.
\begin{table}[htb]
\caption{Payoff tables for the three matrix games.}
\centering
\subfloat[Prisoner's Dilemma]{%
\makebox[0.32\linewidth][c]{%
\begin{tabular}{lcc}
\toprule
 & C & D \\ \midrule
C & $(3,3)$ & $(0,4)$ \\
D & $(4,0)$ & $(1,1)$ \\ \bottomrule
\end{tabular}
}%
\label{tab:ipd}
}
\hfill
\subfloat[Traveler's Dilemma]{%
\makebox[0.32\linewidth][c]{%
\begin{tabular}{lcccc}
\toprule
 & 2 & 3 & $\cdots$ & 10 \\ \midrule
2 & $(2,2)$ & $(4,0)$ & $\cdots$ & $(4,0)$  \\
3 & $(0,4)$ & $(3,3)$ & $\cdots$ & $(5,1)$  \\ 
$\vdots$ & $\vdots$ & $\vdots$ & $\ddots$ & $\vdots$ \\
10 & $(0,4)$ & $(1,5)$ & $\cdots$ & $(10,10)$ \\\bottomrule
\end{tabular}
}%
\label{tab:itd}
}
\hfill
\subfloat[Stag Hunt]{%
\makebox[0.32\linewidth][c]{%
\begin{tabular}{lcc}
\toprule
 & H & F \\ \midrule
H & $(4,4)$ & $(-1,3)$ \\
F & $(3,-1)$ & $(2,2)$ \\ \bottomrule
\end{tabular}
}%
\label{tab:ish}
}
\label{tab:matrix_games}
\end{table}
\paragraph{CleanUp-Lite}
\texttt{CleanUp-Lite} was introduced by \citet{meulemans_multi-agent_2025}. Two agents can move in a $4\times5$ grid, where column $0$ is the \textit{orchard}, and column $3$ is the \textit{river}. Apples can spawn in the orchard, but their spawning probability is inversely proportional to the amount of dirt in the river, which continuously spawns. The dilemma comes from the fact that picking apples gives a reward of $+1$, while cleaning dirt gives no reward at all. Agents also have the option to \textit{zap} the other player for five rounds. For the zap to work the other player has to be in a $2$ cells radius
This environment is particularly interesting because is a mix of a dilemma and a cooperation problem: agents need to learn that cleaning dirt is necessary for apples to grow, but only cleaning gives zero reward, hence, no feedback on their actions. The total reward may be maximized if one agent cleans and one picks apple, but without explicit communication or centralization this is an extremely difficult feat. On the other hand it is relatively simple to learn to just pick apples, but two selfish agents would quickly finish all the apples and saturate the river with dirt. On top of all this, each agent can \textit{directly} punish the other agent by zapping them, which adds the option of choosing an explicitly adversarial behavior.

\subsection{Results: Cooperation in Social Dilemmas}
\label{subsec:coop_results}
We first consider the self-play setting, in which each algorithm plays against a copy of itself. This setting evaluates the ability of agents to cooperate without sharing rewards.
\paragraph{Matrix Games}
Figures~\ref{fig:ipd_welfare}-\ref{fig:stag_welfare}-\ref{fig:trav_welfare} report the social welfare (cumulative reward) during training for the three matrix games for all algorithms.
It is clear from these results that RQA-MASAC outperforms all other algorithms, both in terms of convergence speed, and in quality of the solutions. In all three games our algorithm quickly reaches the optimal social welfare, and it is able to maintain that strategy for the rest of the training, across the different seeds we tested. Interestingly enough, QMIX fails to reach the optimal solution in all three, even having access to the cumulative reward by construction.

\paragraph{Matrix Games with Asymmetric Payoffs}
We also considered the same matrix games, but with asymmetric payoffs between agents. Specifically, one agent keeps the values in Tabs.~\ref{tab:matrix_games}, while the second is modified: for the Prisoner's Dilemma it got $9$ for exploitation, $7$ for mutual cooperation, $-2$ for mutual defection, and $-4$ for being exploited; for the Traveler's Dilemma the bonus is $+1$ and the penalty $-3$; and for Stag Hunt it became $4.5$ for hunting together, $0$ for foraging together, $3.5$ for foraging alone, and $-2$ for hunting alone. The results, shown in Fig.~\ref{fig:asym_ipd_welfare}-\ref{fig:asym_stag_welfare}-\ref{fig:asym_trav_welfare}, highlight how our algorithm can reproduce the same performance even with asymmetric rewards between agents.
\begin{figure*}[bt]
    \centering
    \begin{subfigure}[t]{0.32\textwidth}
        \includegraphics[width=\linewidth]{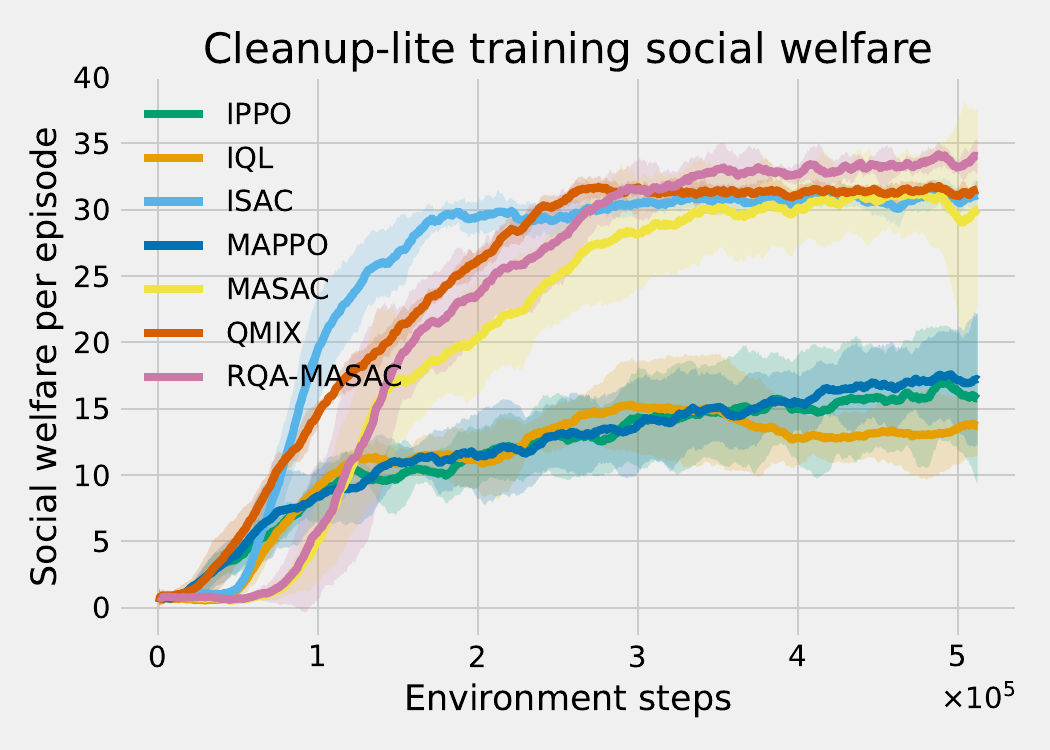}
        \caption{Cumulative social welfare for all algorithms during training on \texttt{CleanUp-Lite}.}
        \label{fig:cleanup_welfare}
    \end{subfigure}
    \hfill
    \begin{subfigure}[t]{0.32\textwidth}
        \includegraphics[width=\linewidth]{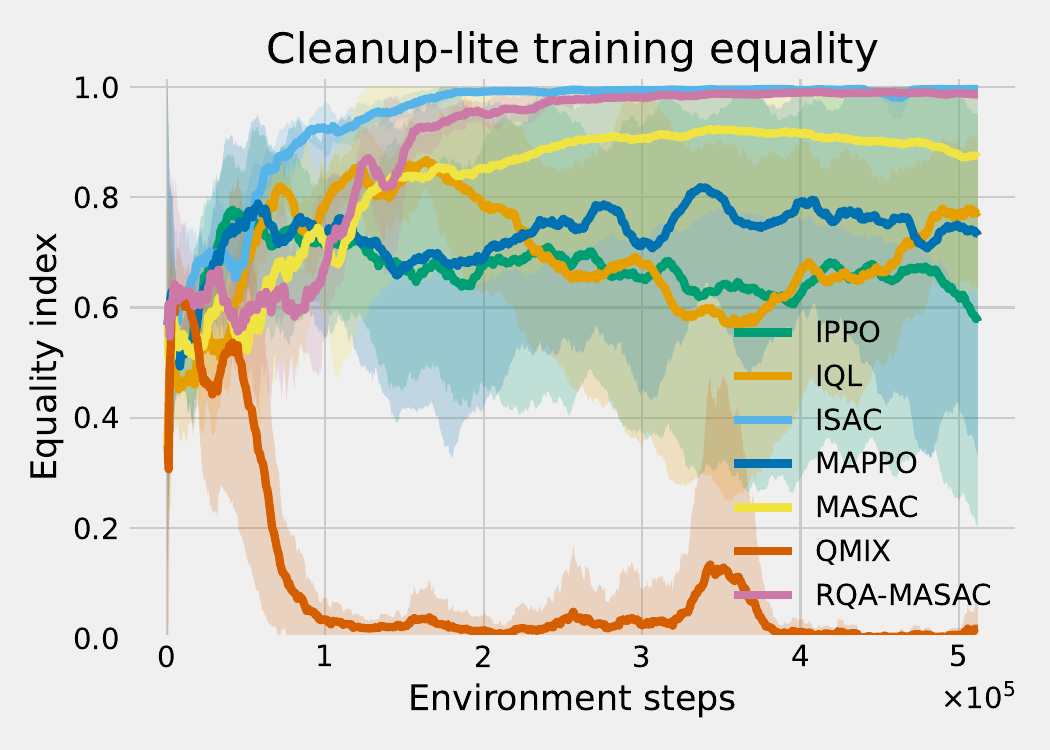}
        \caption{Equality between the two agents during training.}
        \label{fig:cleanup_equality}
    \end{subfigure}
    \hfill
    \begin{subfigure}[t]{0.32\textwidth}
        \includegraphics[width=\linewidth]{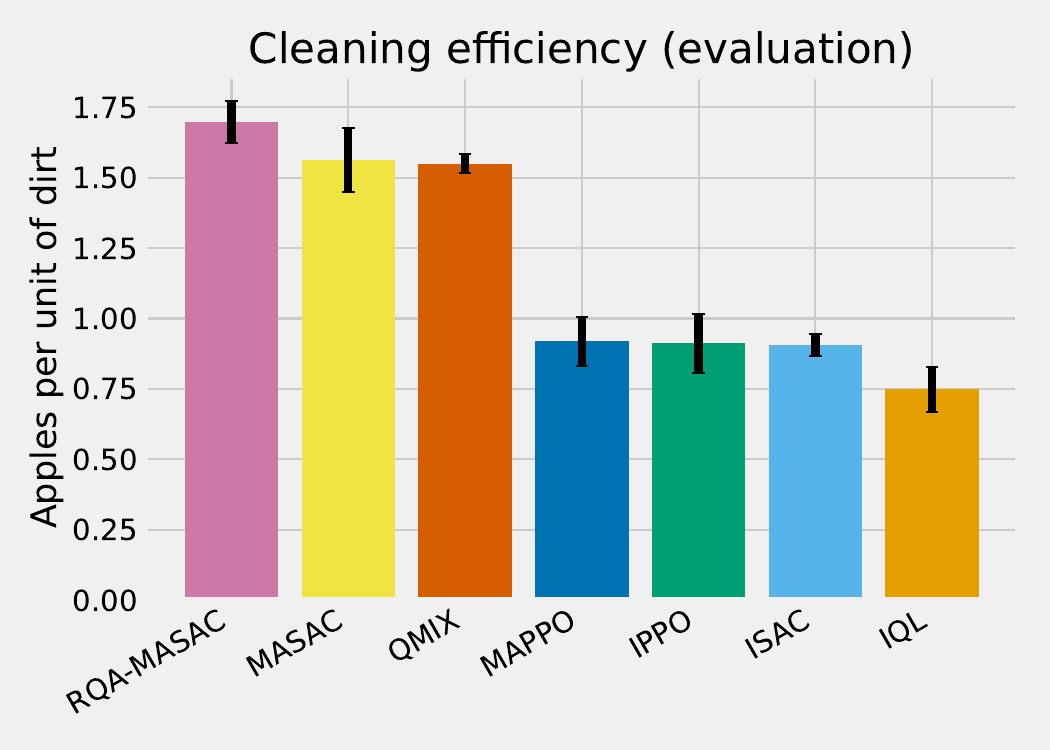}
        \caption{Cleaning efficiency on the evaluation runs.}
        \label{fig:cleanup_efficiency}
    \end{subfigure}
    \caption{(a-b): Plot of the social welfare and equality, averaged across 5 seeds, smoothed over 200 episodes of 64 steps each, and where the shaded area represents the $95\%$ confidence interval. (c): efficiency of each algorithm during evaluation, the error bar represents the standard deviation.}
    \label{fig:cleanup}
\end{figure*}
\paragraph{CleanUp-Lite}
Figure~\ref{fig:cleanup} shows the results on the \texttt{CleanUp-Lite} environment. In Fig.~\ref{fig:cleanup_welfare} we plot the cumulative reward throughout training, and we can see that while ISAC and MASAC can match the results obtained by QMIX, RQA-MASAC still reaches the best result in terms of social welfare. We then compare the equality between agents in Fig.~\ref{fig:cleanup_equality}, computed as $E = 1 - |r_i-r_j|/(r_i+r_j)$: when $E=1$ the two agents are receiving the same reward, $E=0$ one agent is getting the total reward while the other is getting nothing. If the total reward is zero $E$ is undefined, so that step is excluded from the mean. Once again, we see that RQA-MASAC reaches the maximum value, together with ISAC. Interestingly, QMIX, which optimizes the total reward, obtains a high collective score, but at the cost of large reward inequalities between agents. Finally, in Fig.~\ref{fig:cleanup_efficiency} we highlight the efficiency of each algorithm during evaluation, computed as the number of apples picked divided by the amount of dirt cleaned, so that higher values mean higher efficiency. Again, RQA-MASAC is the the most efficient algorithm, meaning that the agents find a better balance between cleaning and picking apples than any other algorithm.
Overall the three results combined show how our modification can be used to obtain high level of cooperation, while maintaining equality between the players and learning an optimal strategy.

\subsection{Analysis: Inducing Cooperation}
\begin{figure}[tb]
    \centering
\includegraphics[width=0.5\linewidth]{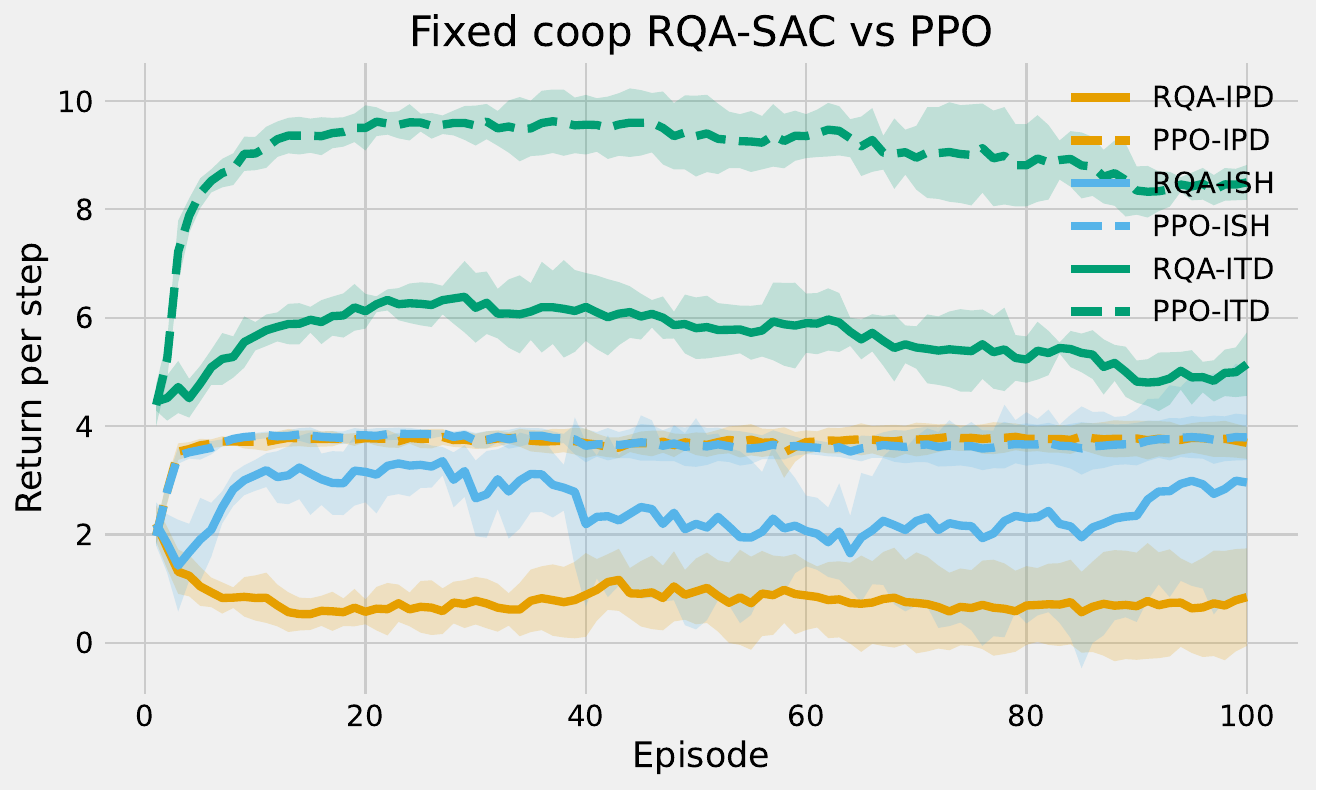}
    \caption{Rewards for the fixed cooperative RQA agent against PPO in the three matrix games. Each line is the average over 10 runs, with the shaded area representing the $95\%$ confidence interval.}
    \label{fig:exploitation}
\end{figure}

When an agent using RQA with only the cooperative augmentation faces an opponent trained with a standard RL algorithm, it is easily exploited. This behavior is illustrated with the Prisoner's Dilemma, Traveler's Dilemma and Stag Hunt in Table~\ref{tab:coop_exploit} and Fig.~\ref{fig:exploitation}, where our agent faces a non-cooperative PPO agent. This is because our agent prioritizes the opponent's welfare at the expense of its own, which a self-interested opponent can take advantage of. This motivates the need of the switching mechanism.
To show that switching between the cooperative and defecting policies can induce cooperation, we evaluate an RQA-MASAC agent against a PPO agent. We choose PPO because both IPPO and MAPPO yielded the lowest social welfare in the previous experiments, which makes it the most challenging opponent to steer toward cooperation. We compare two ways of implementing the switch:
\begin{itemize}
    \item \textit{Frozen model:} we pre-train RQA-MASAC in self-play twice, once with Eq.~\eqref{eq:cooperative_Q} to obtain a cooperative policy and Q-function, and once with Eq.~\eqref{eq:defective_Q} to obtain their defecting counterparts. These frozen policies are then combined through the switching mechanism against a PPO opponent that keeps learning.
    \item \textit{Learning model:} both policies are learned online while playing against the PPO opponent, and the switching mechanism is applied directly during training.
\end{itemize}
We first evaluate RQA on the Iterated Prisoner's Dilemma (IPD), the Iterated Stag Hunt (ISH), and the Iterated Traveler's Dilemma (ITD), where cooperation is clearly defined: mutual cooperation, hunting the stag together, and choosing actions above 8, respectively. We then consider \texttt{CleanUp-Lite}, where cooperation does not necessarily imply an equal split of the reward. To maximize the cumulative reward, agents may specialize, one collecting apples while the other cleans, in which case the cleaning agent receives no direct reward from the environment.

\begin{table}[htb]
\caption{Rewards for fixed cooperative MASAC-RQA (coop-RQA) and PPO, averaged over the last 5 episodes of training.}
\centering
\begin{tabular}{lcc}
\toprule
Game  & coop-RQA & PPO \\ \midrule
IPD  & $0.7 \pm 0.8$ & $3.7 \pm 0.3$ \\
ISH  & $2.9 \pm 1.7$ & $3.8 \pm 0.3$  \\
ITD  & $4.9 \pm 0.3$ & $8.5 \pm 0.2$ \\ \bottomrule
\end{tabular}
\label{tab:coop_exploit}
\end{table}



\begin{figure*}[tb]
    \centering
    \begin{subfigure}[t]{0.48\textwidth}
        \includegraphics[width=\linewidth]{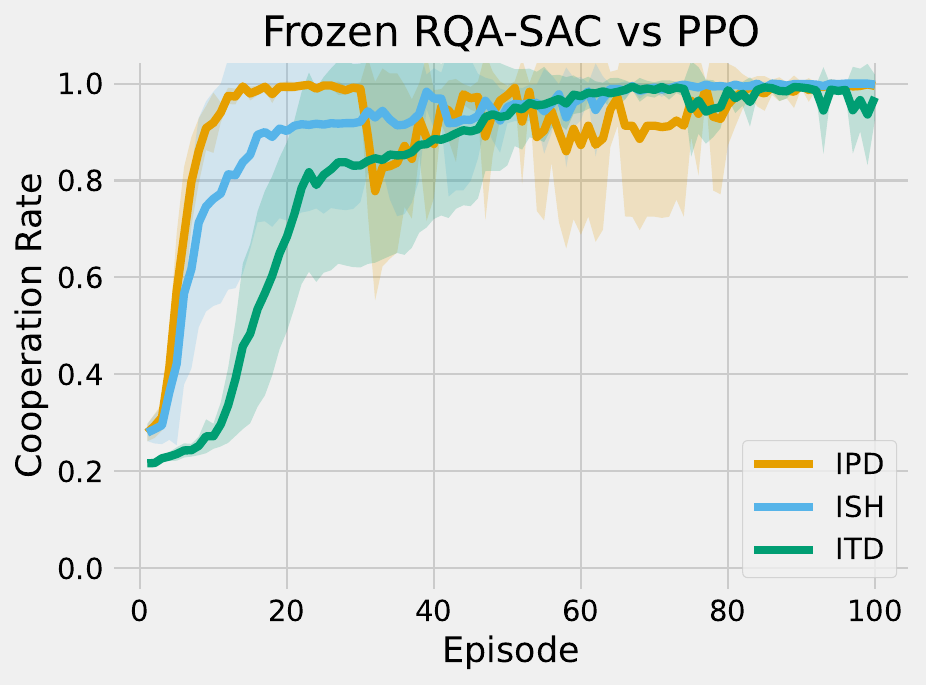}
        \caption{}
        \label{fig:frozen_coop}
    \end{subfigure}
    \hfill
    \begin{subfigure}[t]{0.48\textwidth}
        \includegraphics[width=\linewidth]{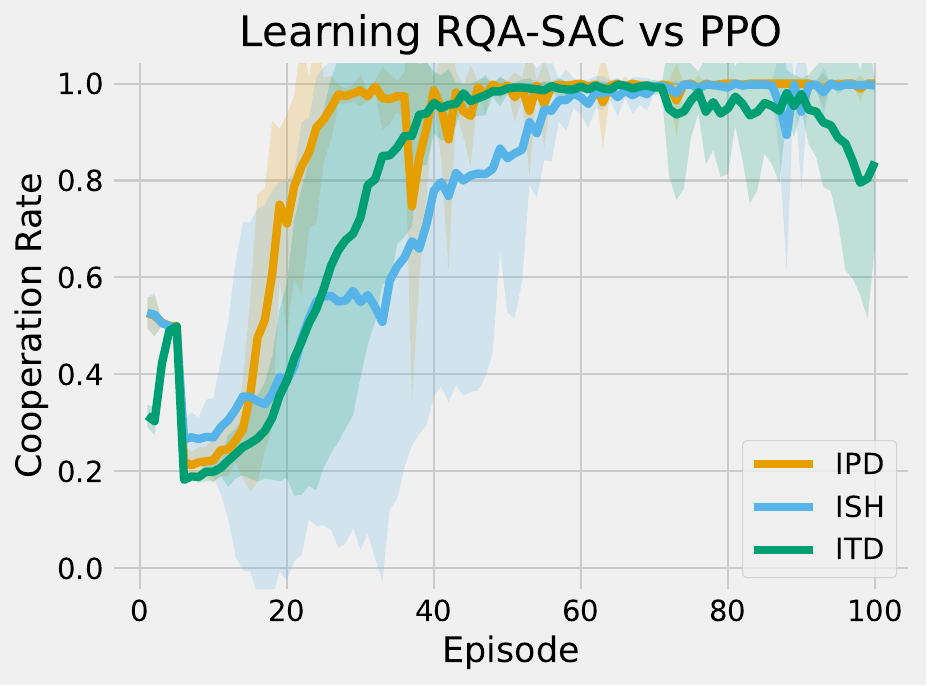}
        \caption{}
        \label{fig:learning_coop}
    \end{subfigure}
    
    \begin{subfigure}[t]{0.48\textwidth}
        \includegraphics[width=\linewidth]{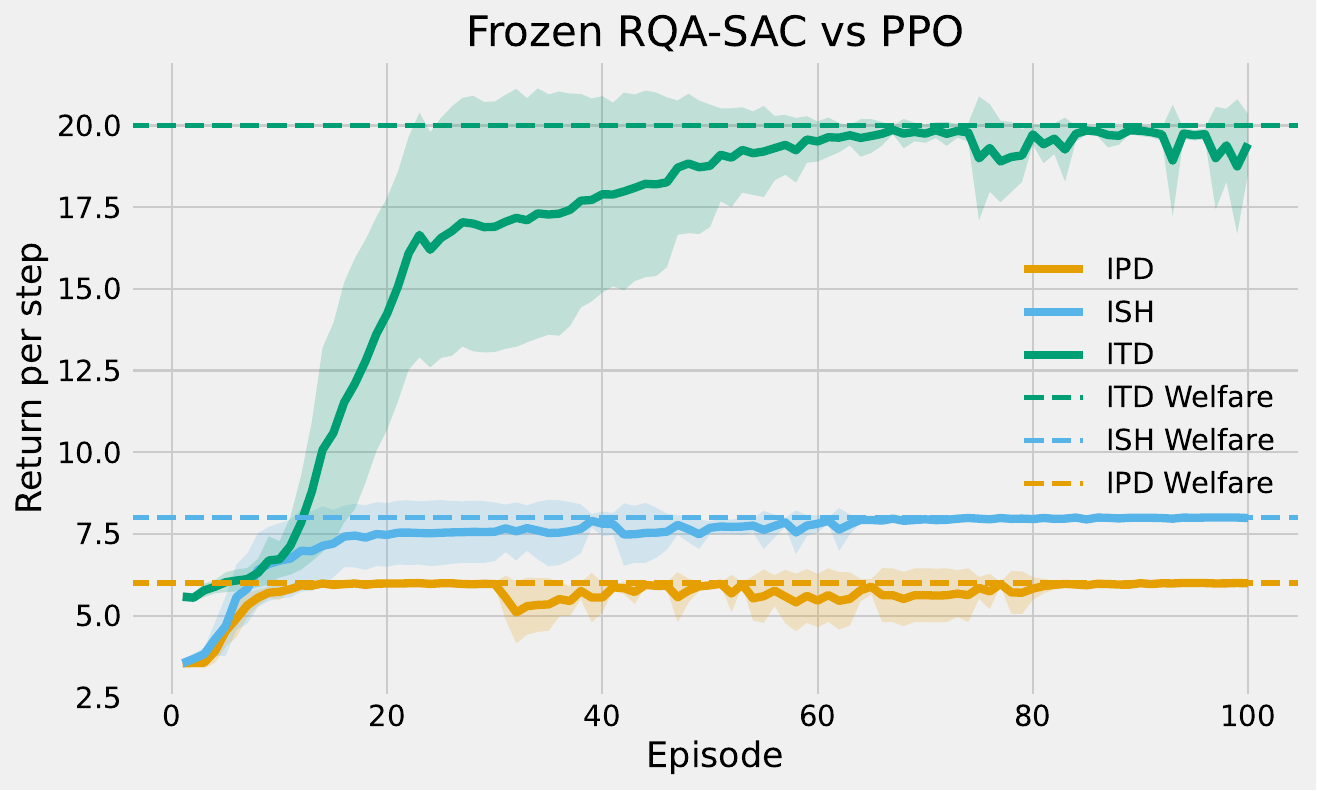}
        \caption{}
        \label{fig:frozen_welfare}
    \end{subfigure}
    \hfill
    \begin{subfigure}[t]{0.48\textwidth}
        \includegraphics[width=\linewidth]{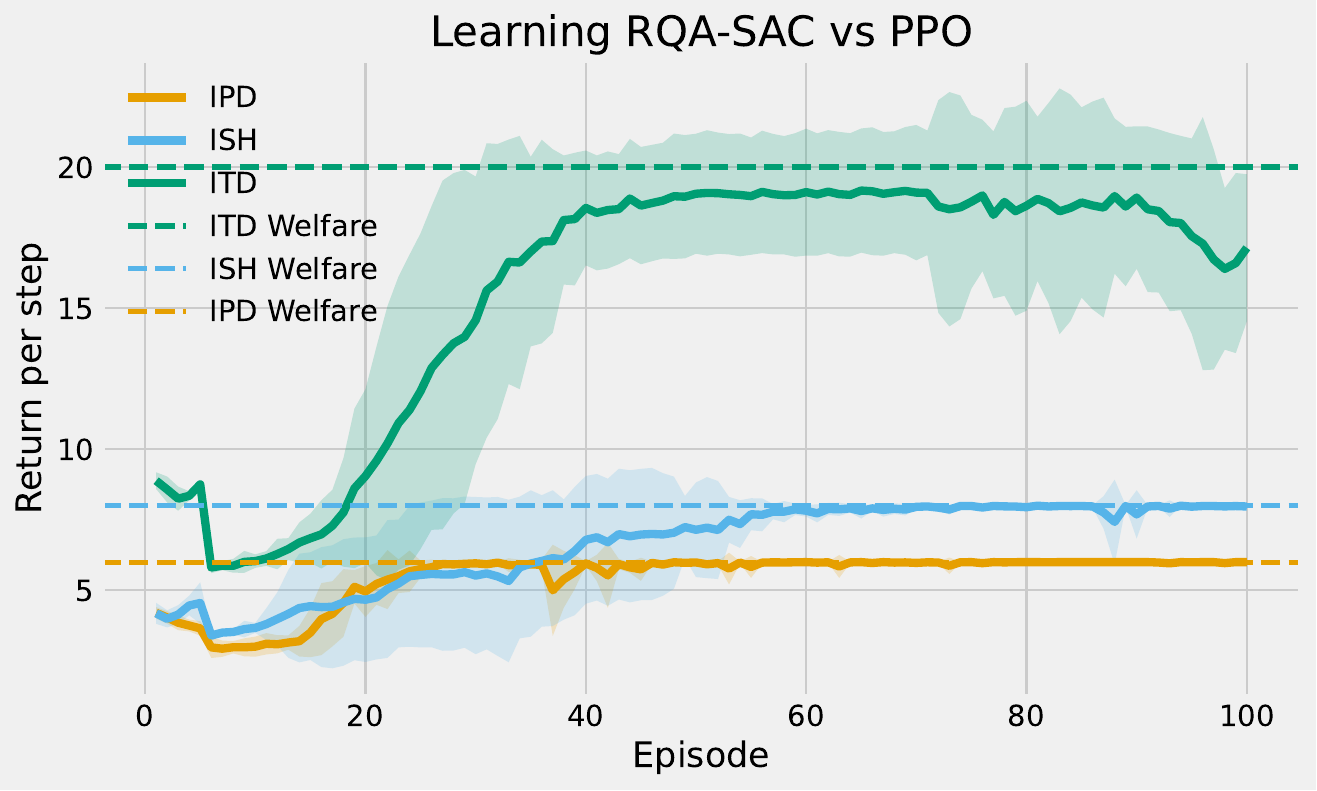}
        \caption{}
        \label{fig:learning_welfare}
    \end{subfigure}
    \caption{(a-c): Plots of the cooperation rate (time passed in cooperative mode by the RQA agent), and the cumulative return for the test with a frozen RQA agent implementing the switching logic against a PPO agent. The dashed lines represent the theoretical maximum welfare. (b-d): Same plots but for a learning RQA agent, implementing the switching logic while training its policies. In both cases and for all matrix games we see that the cooperation rate is very close to 1 by the end of training, and the cumulative reward converges toward the optimal solution.}
    \label{fig:matrix_games_cooperation}
\end{figure*}

\begin{table}[htb]
\caption{Social welfare for PPO under different settings, averaged over the last 5 episodes of training.}
\centering
\begin{tabular}{llccc}
\toprule
Game & Social welfare & self-play & frozen & learning  \\ \midrule
IPD & $6$ & $2 \pm 0$ & $\boldsymbol{5.99 \pm 0.02}$ & $\boldsymbol{5.99 \pm 0.02}$ \\
ISH & $8$ & $4 \pm 0$ & $\boldsymbol{8.00 \pm 0.01}$ & $7.98 \pm 0.04$  \\
ITD & $20$ & $14 \pm 4$ & $\boldsymbol{19.3 \pm 1.7}$ & $16.8 \pm 2.9$  \\ \bottomrule
\end{tabular}
\label{tab:teaching}
\end{table}
\paragraph{Matrix Games}
We report the results in Fig.~\ref{fig:matrix_games_cooperation} and
Table~\ref{tab:teaching}. Specifically in Fig.~\ref{fig:frozen_coop}~-~\ref{fig:frozen_welfare} we report the cooperation rate, computed as the time spent using the cooperative policy by the RQA agent, and the cumulative reward for the \textit{frozen} case, while in Fig.~\ref{fig:learning_coop}~-~\ref{fig:learning_welfare} we report the same results for the \textit{learning} case. In Table~\ref{tab:teaching} we highlight the cumulative reward averaged over the last 5 episodes for both cases, comparing it with the IPPO results from the previous section (under the columns \textit{self-play}), and the optimal social welfare for each game.

Both \textit{frozen} and \textit{learning} improve over self-play, for IPD and ISH both reach the theoretically optimal social welfare (considering the standard deviation). For ITD the \textit{frozen} experiment reaches the optimal solution, while the \textit{learning} one gets close, still improving on the self-play result.
These results support the effectiveness of our switching mechanism. The results of the learning model, which learns both policies online, are particularly significant: without access to the other player's reward, without pre-training, and without explicit communication, our agent induces a high level of cooperation in a co-player trained with PPO, an algorithm that failed to cooperate in the same games in our previous experiment.

\paragraph{CleanUp-Lite}
As discussed previously, \texttt{CleanUp-Lite} differs substantially from the matrix games, as there is no single clearly defined defection strategy. A non-cooperative agent may \textit{zap} the other player or focus exclusively on collecting apples while ignoring cleaning. However, both behaviors can also harm the acting agent: zapping reduces the other agent's contribution to cleaning, while ignoring cleaning allows dirt to accumulate, eventually reducing the number of available apples. Consequently, good performance depends more on effective coordination than solely on avoiding exploitation. For the tests on this environment we considered the \textit{learning} version of the switching agent, with no pre-training.

Table~\ref{tab:zaps} reports attempted and successful zaps by RQA-MASAC under three settings: cooperative RQA-MASAC self-play from Section~\ref{subsec:coop_results}, fixed cooperative RQA (\textit{co-RQA}) against PPO, and switching RQA (\textit{sw-RQA}) against PPO. The switching agent exhibits substantially more aggressive behavior: compared with self-play, attempted zaps increase from $11\times10^3$ to $60\times10^3$, while successful zaps increase from $3.3\times10^3$ to $12\times10^3$. This indicates that the Q-function in Eq.~\eqref{eq:defective_Q} leads the switching agent to adopt zapping as a prominent defection behavior. Zapping also increases for cooperative RQA-MASAC against PPO, suggesting that pairing agents trained with different algorithms can produce a more adversarial interaction than homogeneous self-play.
\begin{table}[ht]
\caption{Total number of zaps during training for RQA-MASAC in different settings, averaged over 5 runs. The values are $\times10^3$.}
\label{tab:zaps}
\centering
\begin{tabular}{l c c c}
\toprule
 & self-play & co-RQA vs PPO & sw-RQA vs PPO \\
\midrule
Attempted & $11 \pm 1.7$ & $31\pm18$ & $\boldsymbol{60\pm29}$  \\
Successful & $3.3 \pm 0.6$ & $6\pm5$ & $\boldsymbol{12\pm5}$ \\
\bottomrule
\end{tabular}
\end{table}
The coordination challenge in \texttt{CleanUp-Lite}, and the limited benefit of relying on zapping alone, are further illustrated in Table~\ref{tab:ppo_stats}. We report apples collected, dirt cleaned, and social welfare for PPO self-play from Section~\ref{subsec:coop_results}, together with \textit{co-RQA vs PPO} and \textit{sw-RQA vs PPO}. Apple and dirt statistics refer to the PPO agent.

Social welfare increases when PPO interacts with either RQA-MASAC variant, with the highest value obtained in the cooperative setting. PPO also collects the most apples in this setting. In contrast, under \textit{sw-RQA vs PPO}, PPO cleans the most dirt, suggesting that switching encourages it to perform more of the unrewarded cleaning task. However, this does not yield the highest social welfare. Under \textit{co-RQA vs PPO}, PPO cleans an amount comparable to \textit{PPO vs PPO}, while social welfare still increases.

These results suggest that when coordination is more important than direct exploitation, as in \texttt{CleanUp-Lite}, the benefit of switching may be limited, while the cooperative variant can achieve higher social welfare by facilitating coordination.

\begin{table}[ht]
\caption{Statistic for the different test with PPO. Values obtained averaging over 5 runs and considering the last 10 episodes of training. Apples and dirt refer to the PPO agent.}
\label{tab:ppo_stats}
\centering
\begin{tabular}{l c c c }
\toprule
 & PPO vs PPO & co-RQA vs PPO & sw-RQA vs PPO\\
\midrule
Apples & $40\pm40$& $\boldsymbol{100\pm60}$& $80\pm29$\\
Dirt &$160\pm40$ &$160\pm90$ & $\boldsymbol{189\pm22}$\\
SW & $14 \pm 5$ & $\boldsymbol{18\pm5}$ & $16\pm2$ \\
\bottomrule
\end{tabular}
\end{table}

\section{Conclusions and Future Work}
\label{sec:conclusion}
In this work, we introduced Reciprocal Q-Augmentation (RQA), a method for promoting cooperation in general-sum multi-agent reinforcement learning without requiring shared rewards, explicit communication, or access to other agents' rewards or policies. RQA uses a latent opponent model to estimate how the other player responds to the agent's actions and incorporates this prediction into the Q-function, allowing the agent to learn both cooperative and defecting behaviors from its own reward.

Our experiments show that cooperative RQA consistently promotes high social welfare across different sequential social dilemmas. In the symmetric and asymmetric matrix games, RQA-MASAC converges more consistently to socially optimal solutions than the evaluated baselines. In \texttt{CleanUp-Lite}, it also achieves high social welfare while maintaining equality between agents and an efficient balance between collecting apples and cleaning dirt.

We further showed that RQA can induce cooperation in an independently learning agent. By switching between cooperative and defecting policies according to the opponent's behavior, RQA-MASAC substantially improves PPO's social welfare in the matrix games, both with pretrained policies and when learning concurrently with PPO. The latter shows that cooperation can emerge without an initial self-play phase, access to the opponent's reward, or explicit coordination between learning algorithms. In \texttt{CleanUp-Lite}, however, the cooperative augmentation achieves higher social welfare than switching, suggesting that reciprocity-based defection is less effective when coordination is the main challenge.

Several directions remain for future work. First, RQA could be extended to settings with more than two agents, scaling perspective taking to account for multiple co-players simultaneously. Second, the switching logic could be learned, allowing the agent to determine when and for how long to switch between its policies, possibly addressing the difference between exploitation-centered environment and coordination-centered ones. These extensions would enable evaluation in larger and more complex multi-agent environments.

\subsection*{Acknowledgements}
This research was enabled in part by support provided by Université de Montréal and the Digital Research Alliance of Canada (\url{https://alliancecan.ca/}).

\bibliography{main}
\bibliographystyle{rlc}

\appendix

\end{document}